\documentclass{bcp_Arch}
\usepackage{graphics}
\usepackage{graphicx}
\usepackage{amsmath}

\def\LaTeX{L\kern -.36em\raise .3ex\hbox{\sc a}\kern -.15em T\kern -.1667em%
\lower .7ex\hbox{E}\kern -.125em X}

\begin{document}

\mathclass{Primary 82C70; Secondary 82C26 and 92C45.}
\thanks{M.M. gratefully acknowledges the support of the German Alexander von Humboldt Foundation through the Fellowship No. IV-SCZ/1119205 STP. The authors are grateful to R. Kouyos, A. Parmeggiani and P. Pierobon for pleasant collaborations on some topics reported in this work.}
\abbrevauthors{M. Mobilia et al.}
\abbrevtitle{Generic principles of active transport}

\title{Generic principles of active transport}

\author{Mauro Mobilia, Tobias Reichenbach, Hauke Hinsch, Thomas Franosch, and Erwin Frey}
\address{Arnold Sommerfeld Center for Theoretical Physics
(ASC) and
  Center for NanoScience (CeNS),\\ Department of Physics,
  Ludwig-Maximilians-Universit\"at M\"unchen,\\ Theresienstrasse 37,
  D-80333 M\"unchen, Germany}

\maketitlebcp

\abstract{Nonequilibrium  collective motion is ubiquitous in nature and often results in a rich collection of intringuing phenomena, such as the formation of shocks or patterns, subdiffusive kinetics, traffic jams, and nonequilibrium phase transitions.
These stochastic many-body features characterize transport processes
in biology, soft condensed matter and, possibly, also in nanoscience.
Inspired by these applications, a wide class of lattice-gas models has
recently been considered.
Building on the celebrated {\it totally asymmetric simple exclusion process} (TASEP)
and a generalization accounting for the exchanges with a reservoir,
we discuss the qualitative and quantitative nonequilibrium properties of
these model systems. We specifically analyze the case of a dimeric lattice gas, the transport in the presence
of pointwise disorder and  along coupled tracks.
}

\section*{1. Introduction.}
The modeling of collective transport phenomena is
an important challenge in theoretical physics, with 
possible technological and interdisciplinary implications (e.g. in biology, soft matter and nanotechnology). 
In this context,  motion of molecular motors [How], single-file diffusion in colloid supsensions 
in confined geometry [Lin], spintronic devices [Zut],
have motivated the study of a class of simple one-dimensional stochastic models.
 Remarkably, and despite their apparent simplicity,
these models have been shown to be characterized by rich physical properties like
nonequilibrium phase transitions, mixed phases, traffic jams, which are
of interest on their own right.
Hence, the goal of this article is to convey a pedagogical review of 
these lattice-gas models and of the techniques which allow to unravel their 
intriguing features.

For the case of biological engines, it is only recently that modern experimental techniques
[Meh] have revealed the detailed causes of sub-cellular motion
and transport. Nowadays, we know that every use of our muscles is the collective effort
of a class of proteins called myosin that ``walk'' on actin filaments.
All proteins that convert the chemical
energy of ATP (adenosine-triphospate) in a hydrolysis reaction into
mechanical work are referred to as molecular motors. These biological ``engines'' are highly
specialized in their tasks: ribosomes
move along mRNA strands while translating the codons into proteins,
dynein is responsible for cilia motion and axonal transport, and
kinesins play a key role in cytoskeletal traffic and spindle formation [How].
Rather than on the exact details of the molecular structure and function of
motor proteins (see e.g. [Schl]), we are interested in phenomena arising out of the
collective interaction of many motors. Early research along this line
was motivated by mRNA translation that is managed by ribosomes, which
 bind to the mRNA strand with one subunit and step
forward codon by codon. To increase the protein synthesis many
ribosomes can be bound to the same mRNA strand simultaneously. This
fact might induce collective properties, the slow down
of an incoming
ribosomes (and of the protein synthesis) due to steric hindrance caused by another ribosome in front, as was first realized by
MacDonald [Mac], who set up a theoretical model for the
translation of highly expressed mRNA.

Much in the spirit of [Mac] our theoretical analysis focuses on simple, yet essential, 
collective processes underlying the motion
and neglect the chemical or mechanical details
on the molecular level of motor steps. Within this approach, even simple stochastic
models are found to exhibit rich collective behavior.

We will start this article by revisiting (Section 2) the main
properties of the celebrated totally asymmetric simple exclusion process
(TASEP). In Section 3,  we review the main
properties of a recently introduced model where the TASEP is coupled
to a process accounting for the exchanges with a reservoir. In the
following sections (Sec. 4-7), we outline recent results  on variants
of such a model and therefore discuss its robustness.  In particular,
the case of a lattice gas of extended particles is considered in Sec.~4, while the effects 
of pointwise disorder is addressed in Sec.~5. Sections 6 and 7 are devoted 
to the transport properties along  two types of coupled tracks. In Section 8, we conclude this work by providing an outlook and our conclusions.

\section*{2. The TASEP.}
Biological engines, like ribosomes and kinesins, often move along one-dimensional tracks (e.g. 
mRNA strands or microtubules).  Each individual ``step'' is the result of a fascinating sequence
of molecular events and is an interesting research topic on its own right [How], but is not our scope here.
In fact, our approach is to devise simple and 
analytically amenable stochastic models. By identifying and mimicking the main ingredients underlying 
the collective transport, our goal is to get collective phenomena which we expect 
to be (qualitatively) comparable to those
observed.

\begin{figure}
\centering
\includegraphics[width=0.5\textwidth]{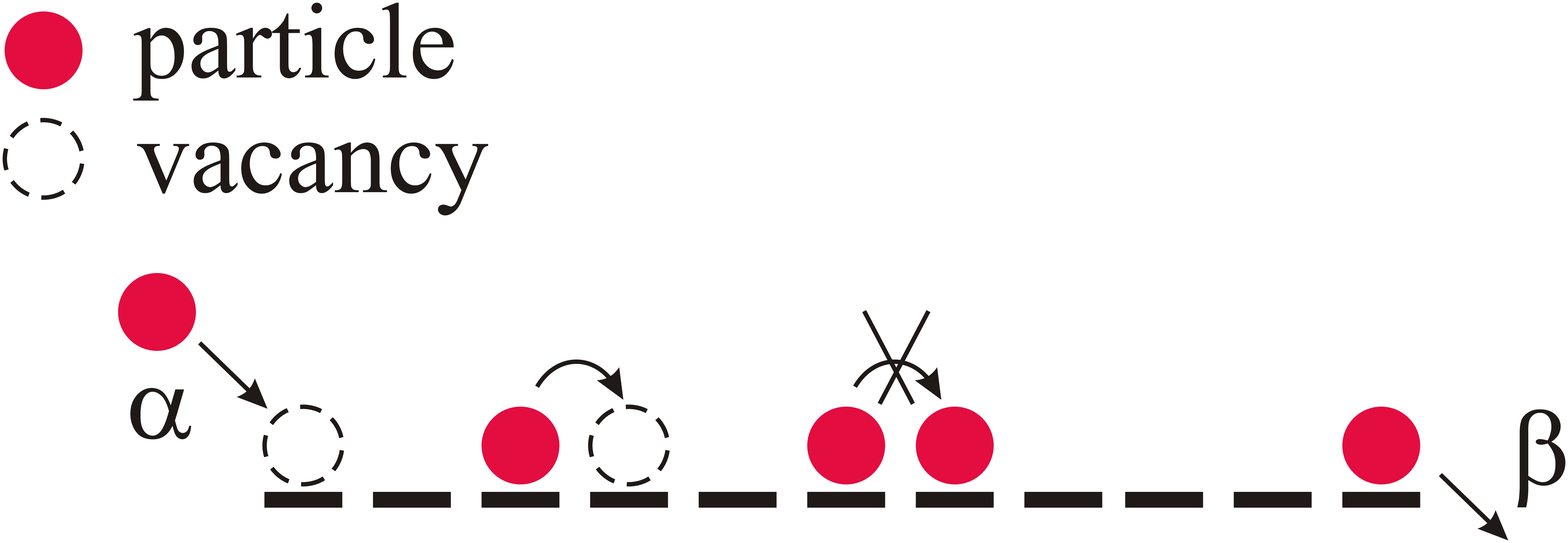}
\includegraphics[width=0.4\textwidth]{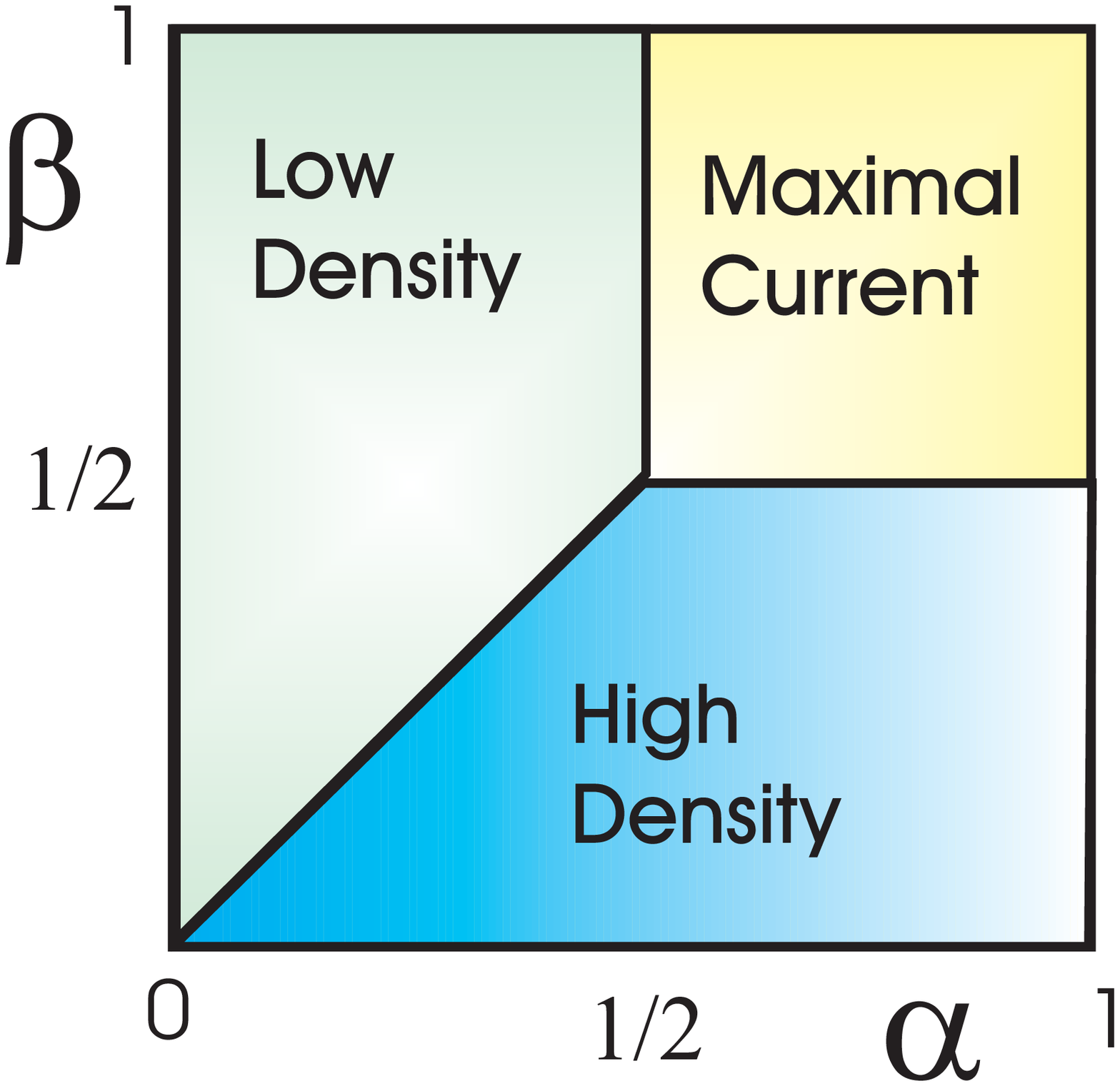}
\caption{{\it(Color online)}. \quad (Left) Schematic model of TASEP (particles are injected with rate $\alpha$, move exclusively to
the right being subject to hard-core exclusion, and are removed with rate $\beta$).
(Right) Phase diagram of TASEP in $\alpha,\beta$
  phase space. One distinguishes the low density (LD, green), the high density (HD, blue) and the maximal
  current (MC, yellow) phases, as well as the first ($\alpha=\beta< 1/2$) and second order
  transition lines (in bold).
}
\label{PhD-TASEP}
\end{figure}

A first step along this line of research was done in a pioneering work by MacDonald et al. [Mac], where the
  the total asymmetric exclusion process (TASEP) was introduced.
  This has since been widely recognized as a paradigmatic model
  in nonequilibrium statistical mechanics. It  consists of a
one-dimensional lattice (Fig. \ref{PhD-TASEP}, left) with $N$ sites labeled
by $i={1,\cdots,N}$ and with a spacing of $a=L/N$. Here, $L$ is the total
length of the lattice and, for technical convenience, is often set to $1$ with
the lattice spacing then referred to as $\varepsilon=1/N$.
Particles (mimicking the motors) have an extension of the size of the
lattice spacing and are subjected to hard core exclusion due to steric
hindrance. Along the track, particles can jump to their right neighboring
site with rate $r$ (set to $1$ in the following) provided the latter is
empty, whence the occupation number $n_i$ of site $i$ takes either  the value 0 or
1.  In addition, one has to specify the dynamic rules at the boundaries of the
lattice. Contrary to equilibrium systems, the latter have a crucial influence
and are responsible for {\it boundary-induced phase transitions} [Kru]. While one can
consider periodic or reflective boundaries, the richest behavior for the TASEP is
obtained in the presence of open boundaries: at the left boundary ($i=1$)
particles attempt to enter the system with a rate $\alpha$, while they exit the lattice at the right boundary
($i=N$) with rate $\beta$. This is equivalent to two additional sites
$i=0$ and $i=N+1$ at the boundaries, which are connected to the system
by the bulk dynamics described above, and are constantly set to the
density $\alpha$ and $1-\beta$ respectively.
Because it does not fulfill the detailed balance, the TASEP evolves
into a steady state where a non-vanishing current is maintained between
boundaries. Upon varying these boundary conditions, TASEP was found to
exhibit phase transitions which -- following general theorems
[Mer] -- are not even allowed for one-dimensional equilibrium
systems in the absence of long-range interactions.

{\it 2.1 Analysis of the TASEP.} While a severe difficulty for the analysis of non-equilibrium systems
stems from the lack of universal concepts like  the Boltzmann-Gibbs
ensemble, models like the TASEP and its generalization can still be studied in many details by means of methods which are
discussed in the remainder of this review article.
Here, for the sake of clarity, we outline the analysis of the TASEP and
some of its main statistical properties. The latter are related to the
density and current distributions, which are quantities of direct biological relevance.

At any given moment, the system can be found in a certain
configuration ${\cal C}$ made up of the occupation numbers at each lattice
site. After a stochastic event occurred, i.e. the jump of one
particle to a neighboring site, the new configuration is ${\cal C}'$. By its
very definition, the TASEP is a Markov process. Hence, the probability of finding the system in the
configuration ${\cal C}$ at time $t$ obeys the
the master equation:
\begin{equation}
\label{ME}
{\rm \frac{d}{dt}}
P({\cal C};t)=\sum_{{\cal C}' \neq {\cal C}}
\left[\omega_{{\cal C}' \to {\cal C}}P({\cal C}';t)-\omega_{{\cal C} \to {\cal C}'}P({\cal C};t)\right] \;,
\end{equation}
where the $\omega_{{\cal C} \to {\cal C}'}$ is the transition rate from the
configuration ${\cal C}$ to ${\cal C}'$.
Using, for instance, the so-called ``quantum
Hamiltonian formalism'' [Sch], one can derive
from (\ref{ME}) the equations of motion for the density
and current of particles. In fact the
density of particle  at site $i$, $\langle n_i\rangle$, is found to obey
\begin{eqnarray}
\label{dens}
{\rm \frac{d}{dt}}\langle n_i\rangle&=&
\langle n_{i-1}(1-n_i)\rangle -\langle n_{i}(1-n_{i+1})\rangle, \quad (i=2,\dots, N-1), \nonumber\\
{\rm \frac{d}{dt}}\langle n_1\rangle&=& \alpha\left[1-\langle n_1\rangle\right]-\langle n_1(1-n_2)\rangle,
\\
{\rm \frac{d}{dt}}\langle n_N\rangle&=& \langle n_{N-1}(1-n_N)\rangle-\beta\langle n_N\rangle,  \nonumber
\end{eqnarray}
where the brackets stand for the average over the histories and, for brevity, we have omitted the
time-dependence. From the discrete continuity equation (with $\nabla  j_i \equiv j_i -  j_{i-1}$), ${\rm \frac{d}{dt}}\langle n_i\rangle+\nabla j_i=0$,
one also infers the expression for the local current of particles:
\begin{eqnarray}
\label{cur}
j_i=
\langle n_i (1-n_{i+1})\rangle
\end{eqnarray}
The interpretation of (\ref{dens},\ref{cur}) is straightforward, but the analysis is difficult because
these equations underlie an infinite hierarchy of equations: the density is connected to the two-point correlations,  which depend on the three-point and so on.
Despite these difficulties, an exact solution has been achieved
by using the so-called matrix and Bethe Ansatz methods [Der1,Der2,Sch].
As these exact methods are mostly tailored for the TASEP, it is insightful to outline an alternative approach. The latter relies on a mean-field assumption and, while it is approximate in its essence, it has a broad range of application and allows to derive the main properties of the models discussed in this review.

We now focus on the stationary properties of the TASEP and show how the phase diagram
can be obtained within a simple  {\it mean-field theory}. The latter relies on
neglecting all the spatial correlations: $\langle \hat n_i \hat n_j \rangle = \varrho_i \varrho_j$, where
we have introduced $\varrho_i\equiv \langle n_{i}\rangle$. In the steady state ($d \varrho_i(t)/dt=0$), this approximation  leads to the following equations of motion for the density in the bulk:
\begin{eqnarray}
\label{densMF}
0=\varrho_{i-1}(1-\varrho_i)-\varrho_i(1-\varrho_{i+1}),
\label{e_steady}
\end{eqnarray}
while the expression for the current reads
\begin{eqnarray}
j_i=\varrho_i(1-\varrho_{i+1}).
\end{eqnarray}
Clearly, the hierarchy of equations is now closed and, together with boundary conditions,
 forms  a nonlinear set of equation giving rise to nontrivial solutions.
 A crucial observation is the local conservation of the stationary current through the bulk.
 In fact, from Eq.~(\ref{densMF}) it follows that $j_i=j_{i-1}$. Actually, one remarks
 that such a relation is also embodied in Eq.~(\ref{dens}) and holds true beyond the
 mean-field approximation. At this point, it is useful to adopt a continuum limit
 which turns the spatial lattice variable continuous. We thus consider a large number $N$ of
lattice sites and rescale the length of the system to $L=1$, with infinitesimal lattice space
$\varepsilon\equiv L/N =1/N\ll 1$. Introducing the spatial variable
$x=i/N$, with $0 \leq x \leq 1$, one can  proceed with a Taylor expansion of the local density:
$
\varrho(x \pm \varepsilon)=\varrho(x) \pm \varepsilon \partial_x
\varrho(x)
+ \frac{1}{2}\varepsilon^2 \partial_x^2 \varrho(x) + O(\varepsilon^3)$.
In the continuum limit and by neglecting the higher order terms in
$\varepsilon$,  Eq.~(\ref{densMF}) is recast as a simple first-order
differential equation:
\begin{equation}
(2 \varrho - 1) \partial_x \varrho  = 0 \;.
\label{e_ode}
\end{equation}
The latter has to be supplemented with the boundary conditions $\varrho(0)=\alpha$ and
$\varrho(1)=1-\beta$.
Because we have a first order differential equation
that needs to satisfy two boundary conditions, we are evidently
concerned with an over-determined boundary value problem. There are
three solutions to (\ref{e_ode}): $\varrho_\mathrm{bulk}(x) = 1/2$
does not satisfy either boundary condition (except for the special
case $\alpha=\beta=1/2$), while $\varrho(x)=C$ can satisfy either the
left or the right boundary condition, resulting in
$\varrho_\mathrm{\alpha}(x)=\alpha$ and
$\varrho_\mathrm{\beta}(x)=1-\beta$, respectively.
To obtain a solution satisfying the boundary conditions, both
solutions need to be matched. To resolve this issue,
both solutions have to be considered valid in non-overlapping areas of the system.
At the point $x_\mathrm{w}$ where the current imposed by the left and right boundaries
coincides (conservation of the current), those areas border and are connected by a sharp domain wall, such that  $\varrho=\varrho_\mathrm{\alpha} $
 for $0 \leq x \leq x_\mathrm{w} \;$ and $\varrho=\varrho_\mathrm{\beta} $
 for $0 \leq x_\mathrm{w} \leq 1 \;$. The domain wall is found to move
 with a velocity $V=\beta-\alpha$ [Kol1]
These simple considerations, together with the conservation of the current, have
important implications and result in three distinct phases :
\begin{itemize}
\item For $\beta<1/2$ and
$\alpha>\beta$, the current
$j_\mathrm{\alpha}=\varrho_\mathrm{\alpha}(1-\varrho_\mathrm{\alpha})$
of the left density solution $\varrho_\mathrm{\alpha}$ is higher than
the current in the right part of the system and the particles are transported
faster to the DW from the
left end, then they head on the right. Hence, the domain wall is
shifted to the left and  the whole
bulk density  takes the value of $\varrho_\mathrm{\beta}$ (except
for a small boundary layer, which vanishes in the limit $N\to \infty$).
This situation therefore corresponds to a  \emph{high density} phase (HD), where both
the density and the current are determined by the exit rate $\beta$: the
mean-field theory gives a spatially constant density $\rho_i=1-\beta$
larger than $1/2$ and a constant current $j_i=\beta(1-\beta)$. In this case,
the current is dominated by the low exit rate which acts as a
bottleneck for the transport.
\item Along the same lines, for $\alpha<1/2$ and $\alpha<\beta$ the low entrance rate
is the limiting factor for the particle current which is now given by
$j_i=\alpha(1-\alpha)$. Since $\rho_i=\alpha$ is always smaller than
$1/2$ in this parameter range, the phase is also termed the \emph{low
  density} (LD) phase.
\item If both $\alpha$ and $\beta$ become larger than the critical value
$1/2$, the current saturates and the density becomes constant $\rho^*=1/2$ independently of the
parameters at the boundaries.
The current is limited by the particle
exclusion in the bulk and its maximal value is $j^*=1/4$; therefore,
this phase was termed \emph{maximal current} (MC) phase.
\end{itemize}

The (first-order) transition line $\alpha=\beta<1/2$ separating the LD and HD phases
 requires special treatment. There, the velocity of the domain wall is zero,
but Monte Carlo simulations and rigorous results [And] show that the DW
actually performs a random walk in a
domain with reflecting boundaries and an average over a sufficiently
long time will therefore result in a homogeneous probability density
over space (instead of a localized shock) $
\varrho(x)=\alpha+(1-\beta-\alpha)x$.

The above discussion is summarized in the phase diagram of the TASEP
(Fig.~\ref{PhD-TASEP}), where the LD and HD phases are separated by a
first-order transition line, while the HD/MC and LD/MC phases
are separated by a second-order transition line. It has to be noticed
that the phase diagram predicted by the mean-field theory coincide with
that obtained by exact methods. Actually, the only features
not accurately accounted for by the mean-field theory are the boundary layers
and the density profile along the line $\alpha=\beta<1/2$ [Der1,Der2,Sch].
The somewhat suprising validity of the mean-field approach can be traced back to the
density relation, $j=\varrho(1-\varrho)$, which has been shown to be an exact result.

\section*{3. The TASEP with on-off kinetics.}
With the aim to cover more realistic situations, 
there has recently been an upsurge of lattice gas models
in the spirit of the TASEP.
In fact, to devise more realistic transport models, some of the simplistic assumptions underlying
the TASEP had to be questioned. As an example, experimental
observations have shown that ribosomes typically cover an area on the
mRNA that exceeds the unit lattice space. To account for
this situation the particles in TASEP have to extend over several
lattice sites. This feature was accounted for by considering the $\ell$-TASEP, where
extended particles (of length $\ell$) move unidirectionally being subject of excluded volume, and it was found
that the TASEP phase diagram was not affected qualitatively [Sha].
It was also shown that in the presence of clusters of bottlenecks, mimicking slow codons
along messenger RNA, may either give rise to local density perturbations or affect
macroscopically the density profile [Chou,Tri].

Intracellular transport along cytoskeletal filaments has also served
as a source of inspiration for driven lattice gas models. While in the TASEP
model motors can only bind and unbind on the left and the right
boundary respectively, cytoskeletal motors are known to detach from the track
to the cytoplasm [How], where they perform Brownian motion, and
subsequently reattach to the track. The interplay between diffusion
in the cytoplasm and directed motion along the filament was studied [Lip]
both in open and closed compartments, focusing on anomalous drift and diffusion
behavior, as well as on maximal current and traffic jams as function of the
motor density. In Ref.~[Par1] it has been realized that,  for an appropriate
scaling of the on-off rates that ensures that particles always travel a finite
fraction on the lattice,  the on-off kinetics may not only
give rise to quantitative changes in the transport efficiency but also to
a novel class of driven lattice gas models. Actually, it was shown that the interplay
between bulk on-off kinetics and driven transport results in a steady state exhibiting phase
separation.  In the following subsection, we review the main
results of the studies [Par1,Par2].

{\it 3.1 The TASEP/LK.}
The crucial feature of cytoskeletal transport is the possibility of
bulk attachment and detachment with a finite residence time on the lattice.
The latter can be understood as an effect of thermal fluctuations that may overcome the
binding energy of the motors that is only of the order of several
$k_\mathrm{B} T$. In this respect, attachment and detachment of motors is a stochastic
process whose dynamic rules have to be defined.  Parmeggiani {\it et
  al.} [Par1] chose to use Langmuir kinetics (LK), known
as adsorption-desorption kinetics of particles on  one- or
two-dimensional lattices coupled to a bulk reservoir [Vil1].
Particles can adsorb at empty sites and desorb from occupied sites, with
microscopic reversibility demanding that the kinetic rates obey detailed
balance leading to an evolution towards an equilibrium  steady state
(in the statistical mechanics sense).
In this framework, the choice of LK is especially appealing as one is then
faced with the competition of a prototypic equilibrium (Langmuir kinetics)
and a paradigmatic non-equilibrium (TASEP) system.  The resulting system -- in the following referred
to as TASEP/LK -- is hence defined as follows: the well-known TASEP is
extended with the possibility of particles to attach to the filament
with rate $\omega_\mathrm{A}$ and to detach from an occupied lattice
site to the reservoir with rate $\omega_\mathrm{D}$.  According to the
type of ensemble (canonical or grand canonical) the reservoir is either
finite or infinite.  Here, the reservoir is assumed to be infinite and
homogeneous throughout space and time. For a lattice
experiencing only LK, the equilibrium coverage depends solely on the ratio
$K=\omega_\mathrm{D}/\omega_\mathrm{A}$, the resulting steady state, given by Langmuir isotherm
$\varrho_\mathrm{L}=K/(K+1)$, is completely uncorrelated
for neglecting any particle interaction
(except excluded volume). This appears sensible as the
diffusion in the cytoplasm is fast enough to flatten any deviations from
the homogeneous reservoir density.
\begin{figure}
\centering
\includegraphics[width=0.7\textwidth]{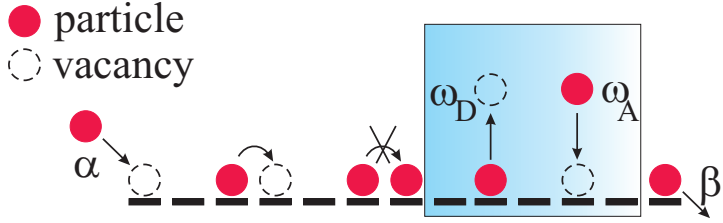}
\caption{Schematic model of TASEP/LK:
  the TASEP is extended by possible particle attachment and detachment
  in the bulk with rate $\omega_\mathrm{A}, \omega_\mathrm{D}$}
\label{f_pff}
\end{figure}
If we now consider the combination of TASEP and LK into the model
displayed in Fig.~\ref{f_pff}, due attention has to be paid to the
different statistical nature of the processes. On one hand, TASEP evolves into a
non-equilibrium state carrying a finite current and, the number of particles being
conserved through the bulk, the system is very sensitive to the boundary
conditions. On the other hand, LK as an equilibrium process is expected to be
robust against any boundary effects (especially for large systems). Combining
both processes on the same time scale would unavoidably lead to a trivial domination
of LK, as the
bulk rates $\omega_\mathrm{A}$ and $\omega_\mathrm{D}$  apply to a
large number of bulk sites and would become predominant over the local entrance and exit rates $\alpha$
and $\beta$.  To observe any interesting behavior (i.e. real interplay)
between the two dynamics, one needs to ensure competition in the system.
A prerequisite for the two processes to compete is to deal with comparable
reaction rates, which requires an appropriate scaling. To this end
$N$-independent global detachment and attachment rates,
$\Omega_\mathrm{D}$ and $\Omega_\mathrm{A}$, are introduced, while the local on-off
rates per site
scale as
\begin{equation}
\omega_\mathrm{D}=\frac{\Omega_\mathrm{D}}{N}; \;
\omega_\mathrm{A}=\frac{\Omega_\mathrm{A}}{N}
.
\label{e_scaling}
\end{equation}
The physical meaning of the above scaling is revealed
by focusing on the involved time scales: a particle on the lattice
performs a move on an time scale which is the inverse
of the move's rate. Therefore, a particle spends an average time
$\approx 1/\omega_\mathrm{D}$ on the lattice before detaching.
Bearing in mind that the TASEP jump rate is set to unity, a particle
 jumps to its right neighboring site typically after one unit time
step. Therefore, the particle will travel a number $N_\mathrm{T} =
1/\omega_\mathrm{D}$ of sites before leaving the lattice. Compared to
the lattice length this corresponds to a fraction of
$n_\mathrm{T}=N_\mathrm{T}/N=1/(N \omega_\mathrm{D})$. To
keep this fraction finite, and ensure effective competition between transport and
on-off dynamics, $\omega_\mathrm{D}$ needs to scale as defined in (\ref{e_scaling}).
Concretely, for the kinesin the ratio between (un)binding and diffusion rates
is typically $10^{-1}-50^{-1}$ [How].

{\it 3.2. Mean Field Solution of TASEP/LK.}
To obtain density and current distributions of the TASEP/LK, we use
again a mean-field approach. Hence, by neglecting any spatial
correlations, along the same lines of Sec.~2.1, the stationary density
is found to obey
\begin{equation}
0 = \varrho_{i-1}(1-\varrho_i)-\varrho_i(1-\varrho_{i+1})-
    \omega_\mathrm{D} \varrho_i + \omega_\mathrm{A} (1-\varrho_i),
\label{e_steady_pff}
\end{equation}
which has to be supplemented with the same boundary conditions as above.
It is again useful to consider the continuum limit, where (at first order in $\epsilon$)
the above equation
reduces to
\begin{equation}
  (2
\varrho - 1)\partial_x \varrho -
\omega_\mathrm{D} \varrho + \omega_\mathrm{A} (1-\varrho)=0, \;
\label{e_ode_pff}
\end{equation}
with the boundary conditions $\varrho(0)=\alpha$ and $\varrho(1)=1-\beta$.
To proceed with the analysis of Eq.~(\ref{e_ode_pff}), it is useful
to introduce the binding constant $K\equiv \omega_\mathrm{A}/\omega_\mathrm{D}$ between the
attachment and detachment rates. However, for the sake of clarity,
here we restrict ourselves to the simple situation with same attachment and detachment rate, i.e. $K=1$,
 and refer the readers to Ref.~[Par2] for a comprehensive discussion of the mathematically more
 involved case where $K\neq 1$. For equal on and off rates, $\Omega_A=\Omega_D=\Omega$,
 it is enlightening to rewrite Eq.~(\ref{e_ode_pff}) as $(\partial_x \varrho - \Omega)(2 \varrho - 1)=0$.
Obviously, there are two solutions to this nonlinear differential equation: the
homogeneous density $\varrho_\mathrm{L}=1/2$, given by the Langmuir
isotherm, and the linear slope $\varrho(x)=\Omega x + C$. The constant
$C$ being determined by the boundary conditions, which leads to one
solution imposed by the entrance boundary, $\varrho_\mathrm{\alpha}(x)=\alpha + \Omega x$, and
another one, namely $\varrho_\mathrm{\beta}(x)=1-\beta-\Omega + \Omega x$,  determined by the exit end.  The complete
density profile $\varrho(x)$ is the combination of one or several pieces of
these three functions.  Depending on how these are matched, we are led
to distinguish several phases as explained below.

\begin{figure}
\centering
\includegraphics[height=7cm]{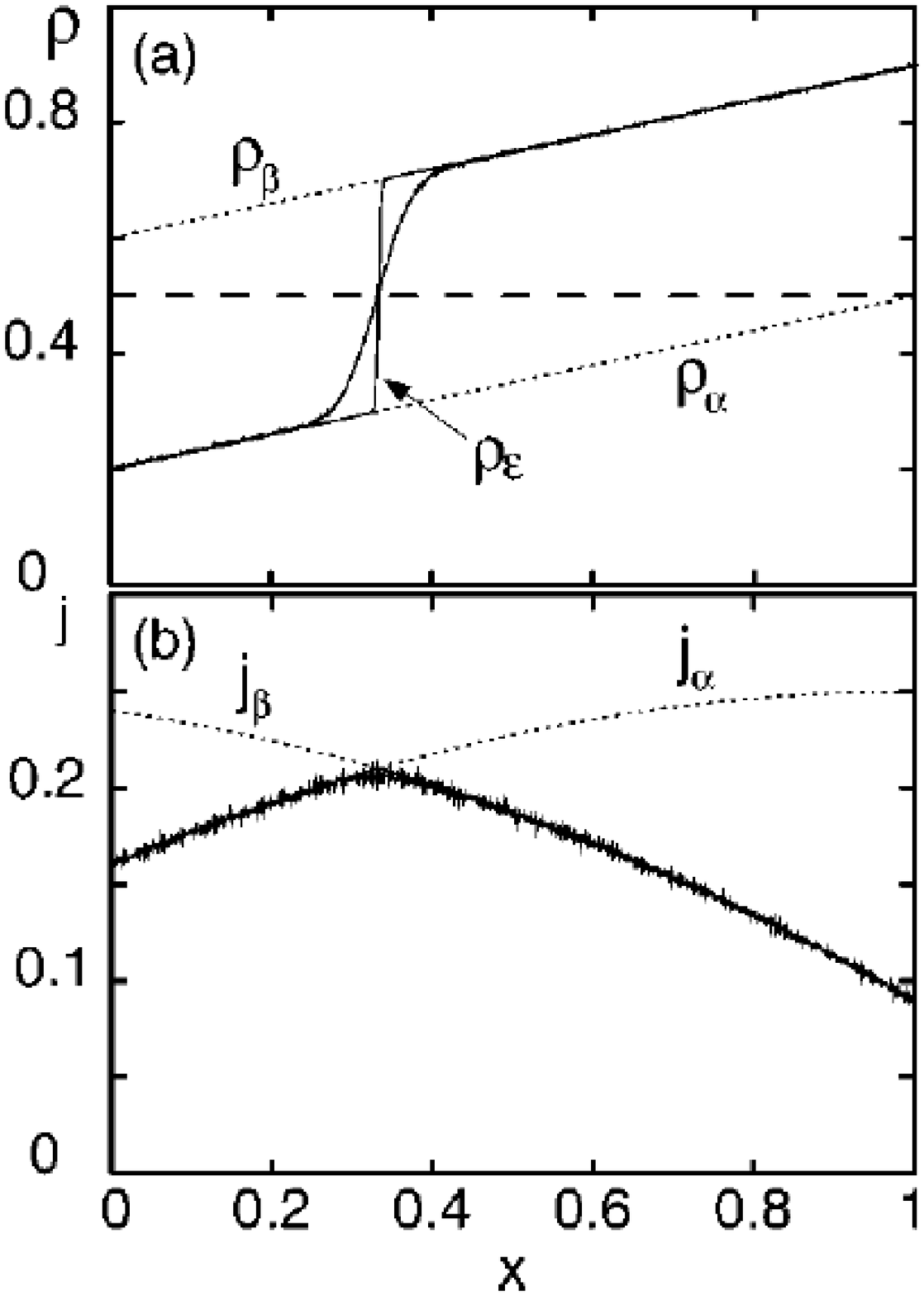}
\hspace{1 cm}%
\includegraphics[height=7cm]{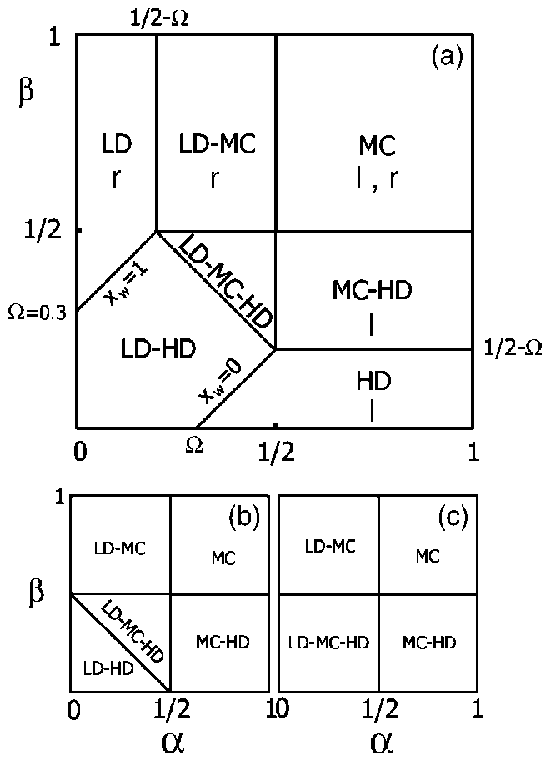}
\caption{Reproduced from [Par2]. \quad(Left) The DW connects the two densities $\rho_\alpha$
  and $\rho_\beta$ ({\it both dashed}) and is localized at the point
  where the correspondent currents $j_\alpha$ and $j_\beta$ match.
  Here low and high densities are connected together
  through a domain wall, and the resulting phase is termed LD-HD.
  Note the finite extend of the domain wall (localization length) that is only
  produced for Monte-Carlo simulations ({\it solid wiggly line}) and
  is not captured by mean-field results ({\it solid line}), reproduced from [Hin1].  
  (Right) Topological changes in the phase diagram of TASEP/LK for
  (a) $\Omega=0.3$, (b) $\Omega=0.5$, (c) $\Omega=1$.}
\label{f_pff_phase}
\end{figure}

{\it 3.3. Phase Diagram and Density Distributions.}
The only region in the parameter space of the TASEP/LK which
is the same as for the simple TASEP is associated with
fast entrance and exit rates. In this case, the system is again in a maximal
current phase, i.e. in a bulk-controlled regime, and the upper right quadrant of the phase diagram (Fig.~\ref{f_pff_phase}, right)
does not change compared to the phase diagram of the TASEP (Fig.~\ref{PhD-TASEP}).
In this regime,  the additional bulk on-off dynamics at Langmuir isotherm
$\varrho_\mathrm{L}=1/2$ does not result in any changes in the density
distribution. In this case, non-equilibrium and equilibrium dynamics
do not compete but cooperate. 

As in TASEP different solutions can be matched in various ways, the
simplest being the connection by a domain wall between the left
solution $\varrho_\mathrm{\alpha}$ and the right solution
$\varrho_\mathrm{\beta}$.  Depending on the current distribution, two
possibilities have to be distinguished. As both solutions are spatially
inhomogeneous, the corresponding currents $j_\mathrm{\alpha}$ and
$j_\mathrm{\beta}$ are strictly monotonic (Fig. \ref{f_pff_phase}, left).
If the currents equal each other inside the system at a
position $x_\mathrm{w}$, the domain wall is localized at this position, as a
displacement to either side would result in a current inequality
driving the DW back to $x_\mathrm{w}$ (see Fig. \ref{f_pff_phase}, left). As a consequence,
the TASEP/LK displays multi-phase coexistence of low and
high density regions (LD-HD phase) in the stationary state on any
time scales. This has to be contrasted with the TASEP,  where this behavior is
only predicted for short observation times. Recently, this domain wall 
localization has been claimed to be observed experimentally [Nish].

If the matching of left and right currents is not possible inside the
system, the known LD and HD phases are found. This is the case for one
boundary condition being considerably larger than the other and depends
 on the slope of the density profile. This slope is determined by the ratio of the TASEP step
rate and the bulk interchange rate $\Omega$: For large $\Omega$ any
density imposed by the boundary relaxes fast against the Langmuir
isotherm of $\varrho_\mathrm{L}=1/2$, resulting  in a steep slope of the
density profile.

This fact allows for the existence of two other phases with
multi-regime coexistence. We could imagine a scenario in which the
boundary imposed density solutions decay fast enough towards the
isotherm to enable a three-regime coexistence of low density, maximal
current and high density (LD-MC-HD phase). Furthermore, a combination
of a MC phase with a boundary layer on one side and a LD or HD region
of finite extend at the other boundary can be imagined. Not all these
phases will be realized for every value of $\Omega$. Instead, the
phase topology of two-dimensional cuts through the
$\alpha,\beta,\Omega$-phase space changes. An example is shown in Fig.
\ref{f_pff_phase} (right).

\section*{4. Dimers and Robustness.}
The results discussed in the previous sections have been obtained
within the framework of generic models. The key notion is that a
detailed description of a physical system is nonessential for the
occurrence of the phenomenon, but a small number of ingredients
encodes a  mechanism that leads to the phenomenon. In order to
corroborate that the models studied so far are representative for
a wide class of models it is instructive to make a step towards
reality and add new features to the model.

\begin{figure}
\centering
\includegraphics[width=0.7\textwidth]{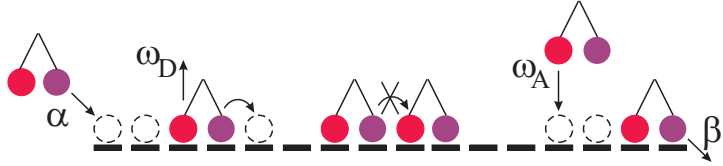}
\caption{(Color online). \quad Illustration of the model of driven dimer transport coupled to a
reservoir} \label{dimers}
\end{figure}
In the case of molecular motors, their molecular structure already
encodes how the motion of a single motors is performed. The most
studied examples are kinesins, dyneins and myosin V which are
known to consist of two sub-components referred  to as \emph{lead}
and \emph{trail head}. These heads bind specifically to a
particular subunit of a long polar molecule acting as molecular
track. As the dynamics of ribosomes along the mRNA motivated the study of $\ell-$TASEP,
 for the transport of molecular motors, it was also suggested that a realistic
 minimal extension of the TASEP/LK
should  comprise \emph{dimers}, i.e. objects that occupy two lattice
sites at any instant of time. These extended objects move according
 the following elementary processes, depicted in Fig. \ref{dimers}, which underlie
 a generalization of the above TASEP/LK dynamics:

\begin{description}
\item[{\it Unidirectional hopping}:] If the lead head of a dimer occupies
a site and
  the following is empty, the dimer advances one step to the right with
  unit rate.

\item[{\it Entrance}:] At rate $\alpha$, a dimer enters the lattice with its trail head
  provided that the first two sites are empty.

\item[{\it Exit}:] A dimer with its lead head on the last site leaves the
lattice with
  rate $\beta$ emptying the last two sites.

\item[{\it Detachment}:] Everywhere in the bulk (i.e.\ trail head on a
site $i=2,\dots,N-2$) a
  dimer leaves the lattice emptying two sites with a site independent
  detachment rate $\omega_D$.

\item[{\it Attachment}:] Everywhere in the bulk a dimer enters the lattice
  with its trail head, provided that the considered site as well as
  its right neighbor is empty, with a site-independent attachment rate
  $\omega_A$.
\end{description}

Although at first glance it appears to be a straightforward
extension of the monomer model, a theoretical description
encounters a number of pitfalls. First, the  mean-field approach
based on ignoring any correlations between neighboring sites is
too simplistic even on the level of pure on-off kinetics. There, the
fact that two neighboring sites have to be empty to allow for an
attachment process modifies the relation of the equilibrium
density $\rho_{\rm L}$, i.e. the Langmuir isotherm, and the binding
constant $K=\omega_A/\omega_D$. Due to the hard-core exclusion,
the corresponding statistical mechanics calculation reduces to a
combinatorial problem~[Tonk,Thom] yielding the
thermodynamic equation of state $\rho_I =
\rho_{\rm L}(K)$~[McG].

The consequences of extended objects on the stationary density
profiles for a \emph{pure} TASEP have been studied in
[Lak,Shaw]. In particular, they found
that the phase diagram is in essence the one of monomeric TASEP, the
differences being of quantitative nature only. A  theoretical
description has been achieved within a 'refined mean-field theory'
that accounts approximately for the correlation induced by
simultaneous hops of two monomers rigidly glued together. Special
care had to be taken to link the density at the boundary to the
entrance/exit rates.

Recently, it has been recognized that the same refined mean-field
ansatz is the proper starting point also to deal with the combined
problem of driven transport (TASEP) and on-off kinetics [Pier1]. The
Langmuir isotherm $\rho_{\rm L}$ as well as an appropriate
current-density relation is encoded in this approach. The density
profile in the stationary state again results from a first order
differential equation subject to two (non-trivial) boundary
conditions.
 In particular, the scenario of phase
separation and phase coexistence has been established for that
model, and it has been rationalized that branch points and domain
walls are generic features of such one-dimensional driven
transport models. An important conclusion of these studies is that
the phenomena that emerge upon weakly coupling a bulk process to a
boundary controlled system are robust.

Stochastic simulations have corroborated the analytically obtained
phase diagram quantitatively. In particular, as a sensitive test
of the theory the position of the domain wall has been measured
and perfect agreement to the theoretical prediction has been
found. Again it appears that a mean-field type approach is able to
capture exactly the phase diagram, as well as the shape of the stationary
density profiles. The weak coupling to the reservoir is
sufficient to wash out the dynamic correlations that build up as
dimers walk through the system. Similarly, the  slow time scale
for the on-off kinetics that intrinsicly would lead to the random
sequential adsorption is sped up by the continuous mass transport
through the system.

\section*{5. Pointwise disorder.}
The presence of some form of randomness on the track is another realistic ingredient which is highly
 desirable to incorporate in a model for intracellular transport.
In fact, {\it  disorder} may biologically be
mediated by structural imperfections of the microtubular structure or
proteins associated to the microtubules that change the affinity of
the motors with the track.
In the TASEP/LK language, imperfections on the microtubule can be mimicked
by the presence of {\it pointwise disorder} [Chou]. In addition, studying
the effect of disorder on dynamic and steady state
properties of nonequilibrium systems is a major challenge in statistical physics [Stin].

\begin{figure}
 \begin{center}
 \includegraphics[width=0.6\columnwidth]{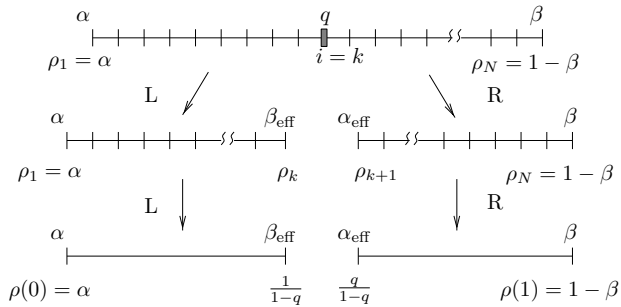}
 \caption{\label{pffdefdec} Reproduced from [Pier2]. \quad
Schematic illustration of the effective mean-field theory: At the defect's location, $i=k$, 
the lattice is split into two subsystems (L and R), which are connected through
the effective rates $\alpha_{\rm eff}$ and $\beta_{\rm eff}$. The last step
represents the continuum limit. The density $\rho(x_d)$ at the bottleneck's location is
$1/(1-q)$ on L and $q/(1-q)$ on R. }
\end{center}
\end{figure}

In this framework,  we have studied the influence of a
bottleneck on the stationary transport properties of the TASEP/LK which is arguably
its simplest, yet nontrivial, disordered version.
Namely, as in the previous sections, we have considered the competition between the totally asymmetric
  exclusion process and Langmuir kinetics in the mesoscopic limit [see Sec.~3.1, Eq.~(\ref{e_scaling})] and in the
presence of open boundaries,  see Fig.~\ref{f_pff}. However, the system is now supplemented with
the presence of a bottleneck of strength $q<1$, which, at a given site $k$, slows down
any incoming particles (Fig.~\ref{pffdefdec}, top). Depending on the current of particles
through the system and the strength of the defect, it is  clear that the latter
may have either a local (spike) or macroscopic effect (jump in the density profile) [Pier2].

In the region of the parameter space where the bottleneck is macroscopically relevant,
the properties of the system are investigated through an {\it effective mean-field theory}
in the continuum limit ($k/N \to x_d$).
As illustrated in Fig.~\ref{pffdefdec}, this approach is built on splitting the lattice into two
subsystems on the left and right of the defect (respectively termed L and R).
To ensure the local conservation of the current (which holds also in the presence of a bottleneck),
one has to couple the L and R subsystems through the effective entrance and exit rates $\alpha_{\rm eff}=q/(1+q)$
and $\beta_{\rm eff}=1/(1+q)$ on the right and left ends of L and R (Fig.~\ref{pffdefdec}).
Therefore, one is left with two coupled TASEP/LK processes whose current profiles have still to be matched. As the current has a spatial dependence (due to the Langmuir kinetics), the defect depletes its profile within a distance $\xi$, that we call \emph{screening length}. As the latter quantity
increases with the strength of the defect and decreases with the attachment-detachment rates,
it is an important concept for our analysis. In fact, depending on the screening length and the position of the defect, one has to distinguish between various scenarios, which are efficiently characterized by the corresponding
 {\it carrying  capacity}. This notion stands for the maximal local current that can flow through the
bulk of the entire system (including the defect).

For the sake of simplicity, here we focus on the situation with equal attachment and detachment rates
(i.e. $\Omega_D=\Omega_A=\Omega$). As the screening length either covers entirely the system
or part of it, we identify four types of
carrying capacities ${\cal C}_i(x), i=1,\dots, 4$.
Using the concepts of screening length and carrying capacity
one readily obtains the phase diagram of the TASEP/LK system perturbed by a localized bottleneck.
In fact, when the boundary currents dominate, the phase behavior of the defect-free system is
recovered. Also, above some critical entrance and exit rates, the
system transports the maximal current (carrying-capacity of the system), independent of the
boundary rates.  Between these two extreme situations, we find
several coexistence phases: above some specific parameter values the system is dominated by the defect and the
phase diagram is characterized by new \emph{bottleneck phases} (BP).
When the screening length spans the whole system, the  associated carrying capacity, say ${\cal C}_1$,
never reaches the maximal value $j^*=1/4$ and one finds {\it four new mixed phases}.
Among the latter, three phases (termed LD-BP, BP-HD and LD-BP-HD, see [Pier2]) are characterized
by localized shocks in the density profile.
When the screening length is short and the related carrying capacity, say ${\cal C}_2$,
saturates at value $j^*$ within both subsystems L and R, one identifies
{\it nine novel bottleneck-induced phases}. When the screening length is
asymmetric (carrying capacities ${\cal C}_3$ and ${\cal C}_4$), i.e. covers 
entirely the subsystem R (L) and partially L (R),
the are {\it six new mixed phases} [Pier2].

\begin{figure}
 \begin{center}
 \includegraphics[width=0.9\columnwidth]{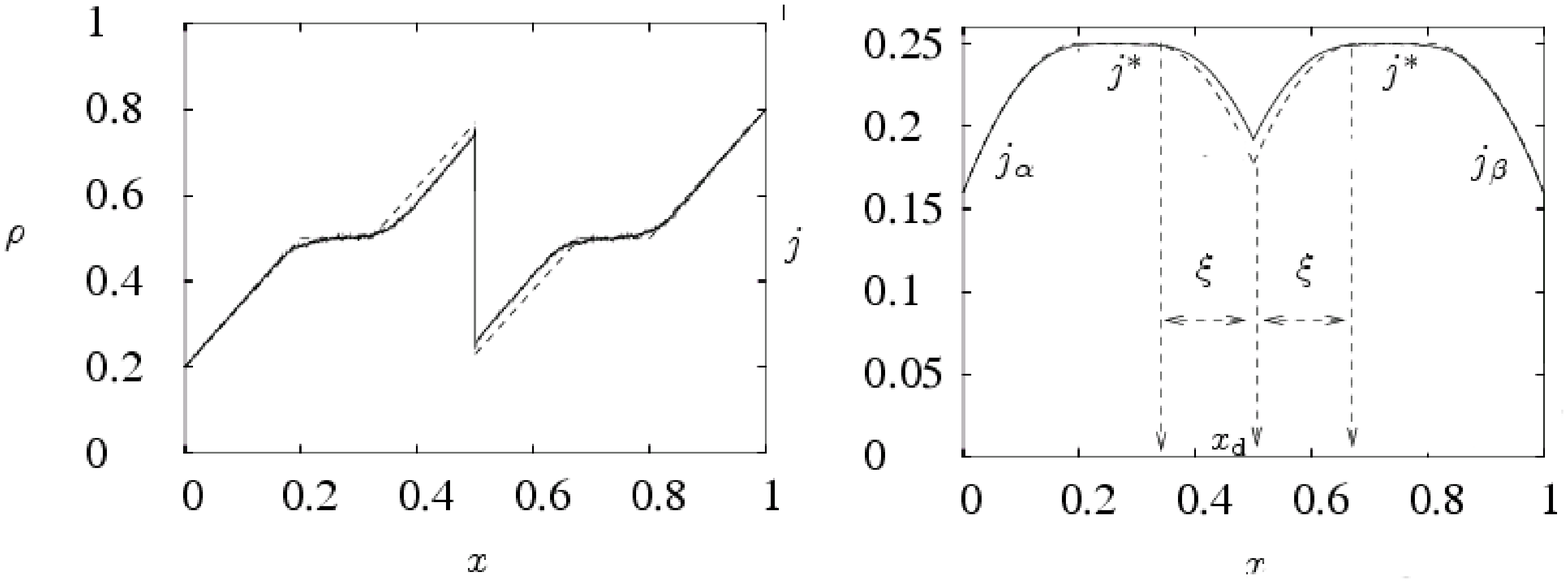}
 \caption{\label{def_profile} 
Example of the TASEP/LK density (left) and current (right) profiles in the presence
  of a bottleneck located at $x_d=1/2$.
    The system (with parameters $q=0.3$, $\alpha=\beta=0.2$, $\Omega_D=\Omega_A=1.5$ and $N=4096$)
 is in a mixed phase (LD-MC-BP-MC-HD), where (from the left to the right end) a low density (LD)
    and maximal current (MC) subphases are followed by a bottleneck-induced
    (sub)phase (BP). The latter is also concatenated to a MC and a high density (HD)
    subphases (see text).  Results of simulations (continuous line) are
    compared to analytic predictions (dashed line). }
\end{center}
\end{figure}

Each of these new mixed phases have been quantitatively studied within our mean-field theory and
the analytical results have been checked against numerical simulations.
As an illustration, let us consider the phase corresponding to the density profile of Fig.~\ref{def_profile}. In this case, the screening length is short and the carrying capacity is of type ${\cal C}_2$. As shown in Fig.~\ref{def_profile} (right), near the extremities of the track,
the current is imposed by the boundaries ($j=j_\alpha$ and $j=j_\beta$) and is below the maximal value $j^{*}$, which is reached within the bulk. This gives rise to a  {\it low} and 
{\it high density} (sub-)phases. At sufficient distance from the ends and from the defect, the current saturates at its maximal value $j^{*}$, which corresponds to a {\it maximal current} (sub-)phase. Within a distance $\xi$ from the defect (screening length; Fig.~\ref{def_profile}, right), the latter dictates the current profile and the system is in a {\it bottleneck phase}.
As a consequence, the density profile displayed in Fig.~\ref{def_profile} (left) corresponds to a phase termed LD-MC-BP-MC-HD.
While the simple TASEP 
is known to be not affected qualitatively  (only some
transitions lines are shifted in the phase diagram) by the presence of a localized defect [Jan,Kol2], 
the effective competition between TASEP and LK
turns out to be very sensitive to the presence of pointwise disorder. 
We expect the same holds for other disordered versions of the TASEP/LK, e.g. 
in the presence of
clusters of bottlenecks [Chou], to which the above effective theory can be adapted.

\section*{6. Multiple parallel lanes.}

\begin{figure}
\begin{center}
\includegraphics[scale=1]{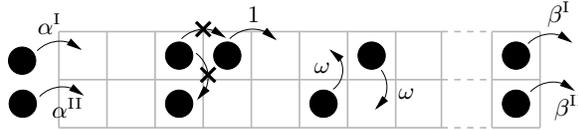}
\caption{Reproduced from [Rei]. Illustration of a two-lane exclusion process. Lane I and II possess individual entering rates,
$\alpha^\text{I}$ resp. $\alpha^\text{II}$ as well as exiting
rates, $\beta^\text{I}$ resp.
$\beta^\text{II}$.\label{cartoon_two_lane}}
\end{center}
\end{figure}

Microtubules, the intracellular tracks for molecular motors like dynein or kinesin [How], are built of 12-14 parallel lanes. Although it has been revealed that the motors proteins typically remain on one track while proceeding on the microtubule, the statistics of deviations (random lane changes) is so far unknown. In the following, we aim to investigate effects of a small, but finite number of lane changes on the resulting density profiles along the lanes by studying a transport model consisting of two parallel lanes I and II, see Fig. \ref{cartoon_two_lane}.

Models with indirect coupling of lanes or strong direct coupling through frequent lane changes have been investigated in [Pop1,Pop2,Pro,Mit], while the focus of the work presented in the following is on weak coupling emerging from only few lane changes. Particles are injected at the left boundary at rates $\alpha^\text{I}$ ($\alpha^\text{II}$) on lane I (II), provided the respective lattice site is empty. In the bulk, particles hop forward at a constant rate which we set to unity; in addition, they may change from lane I to II and back at a small rate $\omega$ (weak lane coupling), always under the constraint of simple site exclusion. Having reached the right boundary, particle extraction occurs at rates $\beta^\text{I}$ ($\beta^\text{II}$).

\begin{figure}[htbp]
\begin{center}
\includegraphics[scale=1]{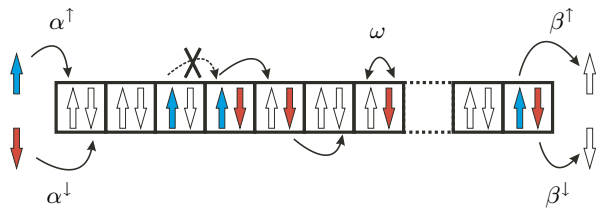}
\caption{(Color online) Reproduced from [Rei]. The same model as above, in the spin-transport interpretation. Particles respect Pauli's exclusion principle; spin flips occur at a (small) rate $\omega$.
\label{cartoon_spin}}
\end{center}
\end{figure}

Interestingly, the above introduced model allows for an interpretation in terms of spin transport. Indeed, on a given lattice site, a particle may be on lane I or II, which may be viewed as an internal state, e.g. the spin state of an electron. Simple site exclusion translates into Pauli's exclusion principle: spin-up and spin-down may simultaneously occupy the same site, though not two particles of the same spin state. The resulting dynamics, depicted in Fig. \ref{cartoon_spin}, may have possible application within the field of spintronics [Zut].

As for the problem of transport with attachment and detachment of hard-core particles in the bulk [Sec.~3.1, Eq.~(\ref{e_scaling})], we employ a mesoscopic scaling for the exchange rate $\omega$
by keeping 
the gross lane change rate, defined as $\Omega=\omega L$, as constant for large system size L. The latter yields a measure for the total number of lane changes that a particle undergoes while traversing the system, which is thus kept fixed, in this way ensuring a competition of lane change events with the boundary processes (injection and extraction).

\begin{figure}[htbp]
\begin{center}
\includegraphics[scale=1]{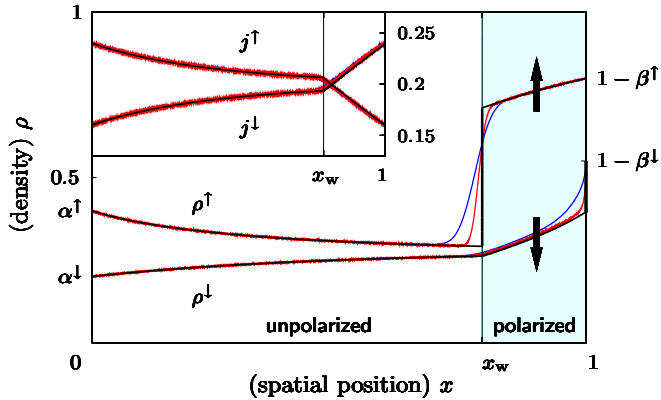}
\caption{Reproduced from [Rei]. Typical scenario for the non-equilibrium stationary state: a shock forms in the spin-up density profile, inducing a spontaneous polarization. The spin currents, shown in the inset, are both continuous. Solid lines correspond to the
  analytical solution, while dashed lines indicate results from stochastic
  simulations on lattices with $L=2000$ (blue or dark grey) and $L=10000$
  (red or light gray).
  \label{EL}}
\end{center}
\end{figure}
This model exhibits a variety of different phases for the emerging non-equilibrium steady state,
controllable by the injection and extraction rates as well as the lane change rate. A typical
scenario is presented in Fig. \ref{EL}, where the language of spin transport has been adopted:
$\rho^\uparrow$ denotes the average density of spin-up (i.e. the concentration of particles on
lane I); and
$\rho^\downarrow$ stands for the density profile of spin-down (lane II) particles. While the density of
spin-down remains at low values, the density of spin-up undergoes a transition at $x_w$ from a region of low to a region of high value with the emergence of a localized shock. In the spin transport
interpretation, we encounter a spontaneous polarization: the densities of both spin states rather equal
each other to the left of the shock, but largely differ to the right, yielding a polarization there.
A key point for the analytical description is the continuity (though not smoothness at $x_w$) of the individual spin currents, shown in the inset.

In the same spirit of the previous sections, we have taken advantage of a mean-field approach and
a continuum limit to analytically describe the stationary state density profiles. Surprisingly,
comparison with stochastic simulations (finite size scaling) uncovers the apparent exactness of the analytical results in the limit of large system sizes. Furthermore, our analytical approach allows for the derivation of phase diagrams, leading to the full knowledge of the system's phase behavior.

\section*{7. SEP and TASEP coupled on a ring.}

\begin{figure}
\begin{center}
\includegraphics[width=8.0cm]{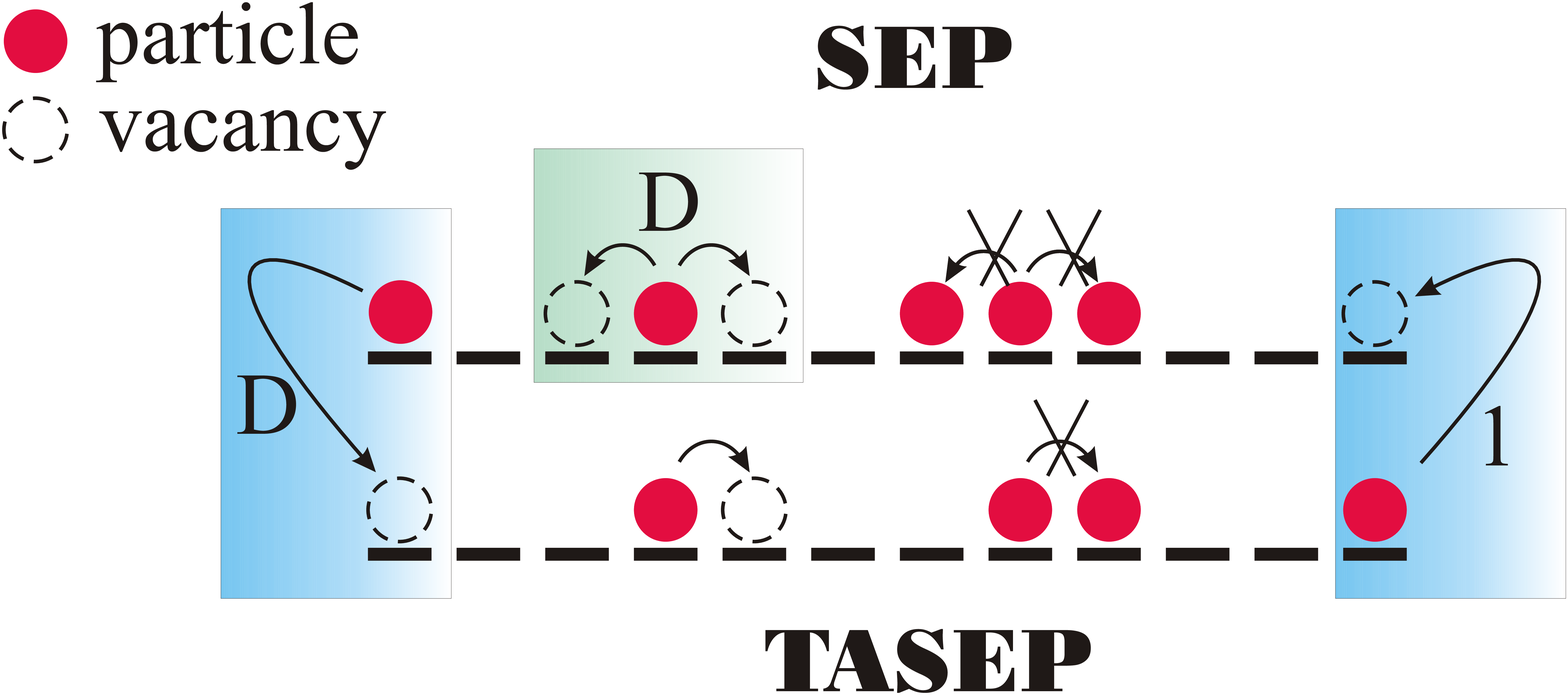}
\includegraphics[width=5.0cm]{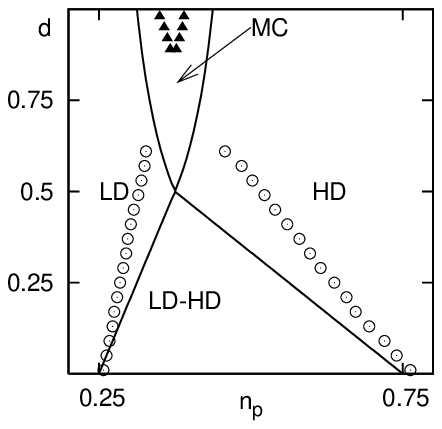}
\caption{({\it left}) (Color online).\quad Schematic illustration of the ring system. The dynamics
  of the upper lane is governed by a SEP with rate D, and the dynamics
  of the lower lane by a TASEP with unity jump rate. The boundary
  densities are indicated at the junctions. ({\it right}) Reproduced from  [Hin2]. Phase
  diagram obtained by MF (solid lines) exhibits four phases.
  Simulations [MC (triangles) and LD-HD (circles) phase boundaries]
  reveal a failure of the MF analysis.}
\label{fig:ring}
\end{center}
\end{figure}

While not only in the TASEP but also in all extended models presented
above, the strategy of a mean-field approach and the continuum limit
has proven succesful to obtain correct phase diagrams and density
profiles in the limit of large system sizes, this cannot a priori be
taken for granted. This was recently demonstrated in a study on the
competitive effects of driven versus diffusive motion in lattice gases
[Hin2]. As prominent examples serve the two paradigms of driven and
diffusive transport in lattice gases: the purely driven TASEP, as
introduced above, and the purely diffusive symmetric exclusion process
(SEP). The analyzed model systems is a periodic setup (Fig.
\ref{fig:ring}, left) consisting of a SEP and a TASEP part of
equal size $N$ with $N_p$ particles inside the system. The hopping
rate of unity on the lower lane signifies a choice of time units and
the hopping rate $D$ on the upper lane fixes the
ratio between driven and diffusive motion. \\
In a first step, we proceed along the lines of analysis used above to
derive a mean-field phase diagram. Here, the central idea is to
decouple the two parts of the ring and consider them as separate lanes
with effective entrance and exit rates. Known results for both
processes can then be applied. While TASEP's non-equilibrium phase
transitions have been introduced in Sec.~2, SEP is known to exhibit a
linear density profile connecting the boundary values [Sch].  The
periodicity of the system has two important consequences: current
conservation and conservation of particle density $n_p=N_p/2N$. The
current conservation allows one to establish the relations $\alpha=D
\delta$ and $\gamma = 1-\beta$ between the boundary densities. The
system's state can then be described by the two remaining control
parameters $D$ and $n_p$. The non-equilibrium steady states have been
classified and a phase
diagram is derived as a function of those two parameters. \\
In the case of TASEP/LK a mesoscopic scaling was used to ensure
competition for arbitrary large systems. To enlight the behavior of
the ring system, the current is of crucial importance. While the lower
lane carries a current $J_L=\rho(1-\rho)$ independent of system size,
the upper lane's current is determined by the boundary condition's
difference as $J_U=(\gamma-\delta)D/N$ reminding of ``Fick's law''.
Obviously, this current decreases with the number of sites, leading to
a ``frozen'' system due to an unphysical vanishing of current.
However, a mesoscopic scaling $d=D/N$ ensures a competition between
both processes and a finite current as expected in the case
of continuous diffusion in the limit $N \to \infty$. \\
Using a mean-field analysis and the conservation laws, phase
transitions can be identified at characteristic changes in the TASEP
density profile. As for the simple TASEP a LD and a HD phase are
observed.  However, a crucial difference is the existence of a
localized domain wall that is observed in simulations and whose
position $x_w$ can be derived as a function of the new control
parameters as $x_{w}=(-3 + 3 d + 4 n_p)/(4 d - 2)$. There exists thus
a LD-HD phase of finite extend in the ring system. This can be
explained by particle conservation and is a consequence of the
periodic system being canonical instead of the grand canonical TASEP.
Additionally, the variance of the domain wall position is found to decay much
faster with $N^{-1}$ compared with a $N^{-1/2}$ decay in TASEP. The
extend of the LD-HD phase is computed to (see Fig. \ref{fig:ring},
{\it right}):
\begin{eqnarray}
d &=& \frac{3-4n_\text{p}}{3} \textnormal{\quad for \quad} x_\text{w}=0 \;, \nonumber \\
d &=& 4 n_\text{p} -1 \textnormal{\quad for \quad} x_\text{w}=1 \;.
\end{eqnarray}
By the introduction of a finite LD-HD phase the first-order phase
transition between LD and HD phase is substituted by two second order
transitions. Similar to TASEP a MC phase can be
identified with phase boundaries:
\begin{equation}
d=\frac{1}{16 n_\text{p} - 4} \qquad \text{and} \qquad d=\frac{1}{8-16 n_\text{p} } \;.
\end{equation}

In a second step, we used Monte Carlo simulations to validate the
derived phase diagram. It is found that strong deviations (symbols in
Fig. \ref{fig:ring} {\it right}) occur. Thinking of the success of the
mean-field method in similar systems, this comes as a surprise at
first sight. Investigating its origin, a closer look at the
density profile reveals that the upper part of the density profile is
actually strictly convex instead of linear. This is a result of a
strong time scale separation of the two subprocesses introduced by the
mesoscopic scaling. While diffusive hopping events happen on a very
short time scale, the entrance of a particle at the r.h.s. junction is
a rather rare event. Between these entrance events the particle
surplus at the r.h.s. relaxes diffusively to the left. As a result the
time averaged density profile appears convex and ``Fick's law'' for
the current is no longer valid. The current is effectively diminished
since it is now bounded to its lowest value at the point of smallest
local gradient. This explains the qualitative shift of the critical
point in the phase diagram to higher $d$. \\
The observed behavior bears analogies to forest fire models [Dro], where out
of two processes of different time scales the faster one strives towards
equilibrium and the slower one occasionally forces the system to a
non-equilibrium state. It furthermore reminds of the coexistence line
in simple TASEP with open boundaries where a linear profile is
obtained as the time average over a randomly walking domain
wall. Hence, the study of the ring system highlights the importance of
time scale separation and points out the limits of mean-field theory
in lattice gas models.

\section*{8. Outlook and conclusion.} In this work, we have described how 
complex nonequilibrium transport phenomena, e.g. the motion
of molecular motors along microtubuli or transport of spins, have inspired stochastic models displaying rich nonequilibrium collective
properties. Namely, in the same spirit as mRNA translation 
was modeled by means of the {\it totally asymmetric simple exclusion process} (TASEP) [Mac], 
intracellular transport processes have motivated the 
study of a broad class of lattice gas models. To account for exchanges with a reservoir (e.g. mimicking
 the binding/unbinding of motors from/to the cytosol) and the track, 
 TASEP was supplemented with an on-off (Langmuir) kinetics.  
 An effective competition between driven motion and Langmuir kinetics arises
 provided  the related attachment and detachment rates scale with the inverse of the system size.
 In this case,  a finite residence time on the lattice is ensured and results in the emergence of 
 a new class of non-equilibrium features, like multi-phase coexistence, localized shocks and enriched phase diagrams.
Along these lines, we have reported recent results for physically motivated variants of the above
model.

Specifically, we have discussed transport properties of a one-dimensional gas of dimers, which was
motivated by the fact that molecular motors like kinesin have two heads. The finite extension of the particles then has led to a
breaking of particle/hole symmetry accompanied by a failure of a
na{\"i}ve mean-field approach. A more sophisticated approach had
to be developed that  allows to calculate the exact phase diagram
(essentially) analytically. It has been argued that the scenario
of phase separation is generic for this kind of models. 
We have also considered the effects of pointwise disorder in its simplest form, i.e.
 in the presence of a localized  {\it bottleneck} slowing down any incoming particle. 
By means of an effective mean-field theory and numerical simulations, we have shown that such a single defect is responsible for the emergence of numerous new mixed phases and can result in the simultaneous formation of two traffic jams.
The presence of multiple parallel lanes on which motors proceed can lead
to further interesting phenomena.  We have studied a two-lane model with rare lane 
switch events of the particles and discussed
the mapping to a possible spintronic device.
Depending on boundary rates, the stationary density profiles can exhibit jammed states
appearing on one or both lanes. Finally, the study of competition between driven and diffusive motion on two lanes in a periodic setup has shown the limitations of mean-field theory in the presence of strong time-scale separation.

\references{And}
{
\item{[And]} E. D. Andjel, M. D. Bramson, and T. M.  Liggett, {\it Shocks in the asymmetric exclusion process} Probab. Th. Rel. Fields {\bf 78} 231 (1988).
\item{[Chou]} T. Chou and G. Lakatos, {\it Clustered Bottlenecks in mRNA Translation and Protein Synthesis},
 Phys. Rev. Lett. {\bf 93}, 198101 (2004)
\item{[Der1]} B. Derrida, E. Domany, D. Mukamel, {\it An exact solution of a one-dimensional asymmetric exclusion model with open boundaries}, J. Stat. Phys.
  \textbf{69}, 667 (1992).
\item{[Der2]} B. Derrida, M.R. Evans, {\it  Exact correlation-functions in an asymmetric exclusion model with open boundaries}, J. Physique I \textbf{3}, 311 (1993).
\item{[Dro]} B. Drossel and F. Schwabl, {\it self-organized critical forest-fire model},
 Phys. Rev. Lett. {\bf 69}  (1992), 1629.
\item{[Gar]} C. W. Gardiner, {\it Handbook of Stochastic Methods}, 1st ed., Springer, Berlin, 1983.
\item{[Hin1]} H. Hinsch, R. Kouyos and E. Frey, {\it From intracellular traffic to a novel class of driven lattice gas models} in {\it Traffic and Granular Flow '05}, edited by A. Schadschneider, T. P\"oschel, R. K\"uhne, M. Schreckenberg, and D. E. Wolf (Springer, Berlin, 2006).
\item{[Hin2]} H. Hinsch and E. Frey, {Bulk-Driven Nonequilibrium Phase Transitions in a Mesoscopic Ring} , Phys. Rev. Lett. {\bf 97} (2006), 095701.
\item{[How]} J. Howard, {\it Mechanics of Motor Proteins and the Cytoskeleton}, Sinauer Press, Sunderland, Massachusetts, 2001.
\item{[Jan]} S. A. Janowsky and J. L. Lebowitz, Phys. Rev. A {\bf 45}, 618 (1992).
\item{[Kol1]} A.B. Kolomeisky, G.M. Sch\"utz, E.B.
  Kolomeisky, J.P. Straley, {\it Phase diagram of one-dimensional driven 
  lattice gases with open boundaries}, J. Phys. A \textbf{31}, 6911 (1998).
\item{[Kol2]} A. B. Kolomeisky, {\it Asymmetric simple exclusion model with local inhomogeneity}, J. Phys. A {\bf 31}, 1152 (1998).
\item{[Kru]} J. Krug, {\it Boundary-induced phase transitions in driven diffusive systems}, Phys. Rev. Lett. {\bf 67}, 1882 (1991).
\item{[Lak]} G. Lakatos and T. Chou, {\it Totally asymmetric exclusion processes with particles of arbitrary size},  J. Phys. A: Math. Gen. {\bf 36}, 2027 (2003).
\item{[Lin]} B. Lin, M. Meron, B. Cui, and S. A. Rice,
 {\it From Random Walk to Single-File Diffusion},  Phys. Rev. Lett. {\bf 94}, 216001 (2005) 
\item{[Lip]} R. Lipowsky, S. Klumpp, T.M. Nieuwenhuizen,
 {\it Random Walks of Cytoskeletal Motors in Open and Closed Compartments}, Phys. Rev.  Lett. \textbf{87}108101 (2001).
\item{[Mac]} C.T. MacDonald, J.H. Gibbs and A.C.Pipkin, {\it 
Kinetics of biopolymerization on nucleic acid templates}, Biopolymers \textbf{6}, 1 (1968).
 \item{[McG]} J. McGhee and P. von Hippel, {\it Theoretical aspects of DNA-protein interactions: Co-operative and non-co-operative binding of large ligands to a one-dimensional homogeneous lattice}, J. Mol. Biol. 86, 469 (1974).
\item{[Meh]} A.D. Mehta, M. Rief, J.A. Spudich, D.A. Smith, R.M.
 Simmons, {\it Single-molecule biomechanics with optical methods },
 Science \textbf{283}, 1689 (1999).
\item{[Mer]} N.D. Mermin, H. Wagner, {\it Absence of Ferromagnetism or Antiferromagnetism in One- or Two-Dimensional Isotropic Heisenberg Models}, Phys. Rev. Lett. \textbf{17},
  1133 (1966).
\item{[Mit]} T. Mitsudo and H. Hayakawa, {\it Synchronisation of kinks in the two-lane totally asymmetric simple   exclusion process with open boundary conditions}, J. Phys. A: Math. Gen., {\bf 38}, 3087 (2005). 
\item{[Nish]}  K. Nishinari, Y. Okada, A. Schadschneider, D.
  Chowdhury, {\it Intracellular Transport of Single-Headed Molecular Motors KIF1A}, 
  Phys. Rev. Lett \textbf{95}, 118101 (2005).
\item{[Par1]} A. Parmeggiani, T. Franosch and E. Frey,  {\it Phase Coexistence in Driven One Dimensional Transport}, Phys. Rev. Lett. 90, 086601 (2003).
\item{[Par2]}A. Parmeggiani, T. Franosch and E. Frey, {\it Totally asymmetric simple exclusion process with Langmuir kinetics  } Phys. Rev. E {\bf 70}, 046101 (2004).
\item{[Pier1]}P. Pierobon, T. Franosch, and E. Frey. {\it Driven lattice gas of dimers coupled to a bulk reservoir}, Phys. Rev. E {\bf 74}, 031920 (2006).
\item{[Pier2]} P. Pierobon, M. Mobilia, R. Kouyos, and E. Frey, {\it Bottleneck-induced transitions in a minimal model for intracellular transport}, Phys. Rev. E {\bf 74}, 031906 (2006).
\item{[Pop1]} V. Popkov and I. Peschel, {\it Symmetry breaking and phase coexistence in a driven diffusive   two-channel system}, Phys. Rev. E {\bf 64}, 026126 (2001). 
\item{[Pop2]} V. Popkov and G. M. Sch\"utz, {\it Shocks and excitation dynamics in a driven diffusive two-channel system}, J. Stat. Phys {\bf 112}, 523 (2003).
\item{[Pro]}  E. Pronina and A. B. Kolomeisky, {\it Asymmetric Coupling in Two-Channel Simple Exclusion Processes} Physica A {\bf 372}, 12 (2006). 
\item{[Rei]} T. Reichenbach, T. Franosch, and E. Frey, {\it Exclusion Processes with Internal States},  Phys. Rev. Lett. {\bf 97}  (2006), 050603.Stinchcombe R
\item{[Sch]} G. Sch\"utz in {\it Phase Transitions and Critical Phenomena},
edited by C. Domb and J. Lebowitz (Academic Press,
London, 2000).
\item{[Schl]} M. Schliwa and G. Woehlke, {\it Molecular motors }, Nature \textbf{422}, 759  (2003).
\item{[Sha]} L.B. Shaw, P.K.P. Zia, K.H. Lee, {\it Totally asymmetric exclusion process with extended objects: A model for protein synthesis}, Phys. Rev. E  \textbf{68}, 021910 (2003).
\item{[Stin]} R. B. Stinchcombe, {\it Disorder in non-equilibrium models }, J. Phys.: Condens. Matter. {\bf 14}, 1473 (2002).
\item{[Thom]}  C. Thompson, {\it Classical Equilibrium Statistical Mechanics} (Oxford University Press, New York, 1988).
\item{[Tonk]} L. Tonks, {\it The Complete Equation of State of One, Two and Three-Dimensional Gases of Hard Elastic Spheres}, Phys. Rev. {\bf 50}, 955 (1936).
\item{[Tri]}G. Tripathy and M. Barma, {\it Driven lattice gases with quenched disorder: Exact results and different macroscopic regimes}, Phys. Rev. E \textbf{58}, 1911 (1998)
\item{[Vil1]} E. Frey, A. Vilfan, {\it Anomalous relaxation kinetics of biological lattice-ligand binding models}, Chem. Phys.
  \textbf{284}, 287 (2002).
\item{[Van]} N.G. van Kampen, \emph{Stochastic Processes in
    Physics and Chemistry} (North Holland, Amsterdam, 1981).
\item{[Zut]} $\check{Z}$uti$\acute{\text{c}}$, J. Fabian, and S. Das Sarma, {\it Spintronics: Fundamentals and Applications},  Rev. Mod. Phys.  {\bf 76} (2004), 323-410.
}

\end{document}